\documentclass[preprint]{aastex}
\usepackage{graphicx}

\shorttitle{Comments on Monoceros}
\shortauthors{L\'opez-Corredoira et al.}

\begin{document}

\title{Comments on the ``Monoceros'' affair}

\author{M. L\'opez-Corredoira\altaffilmark{1,2}, A. Moitinho\altaffilmark{3},
S. Zaggia\altaffilmark{4}, Y. Momany\altaffilmark{5}, G. Carraro\altaffilmark{5}, 
P. L. Hammersley\altaffilmark{6}, A. Cabrera-Lavers\altaffilmark{1,2}, 
\& R. A. V\'azquez\altaffilmark{7}}
\affil{$^1$ Instituto de Astrof\'\i sica de Canarias, E-38200 La Laguna, 
Tenerife, Spain\\
$^2$ Departamento de Astrof\'\i sica, Universidad de la Laguna, E-38206 La Laguna, 
Tenerife, Spain\\
$^3$ SIM, Faculdade de Ci\^encias da Universidade de Lisboa, Ed. C8. Campo Grande 1749-016 
Lisbon, Portugal\\
$^4$ Istituto Nazionale di Astrofisica, Osservatorio Astronomico di Padova, Vicolo dell'Osservatorio 5, 35122, Padova, Italy\\
$^5$ European Southern Observatory, Alonso de Cordova 3107, Vitacura, Santiago, Chile\\
$^6$ European Southern Observatory, Karl-Schwarzschild-Strasse 2, Garching, D-85748, Germany\\
$^7$ Facultad de Ciencias Astron\'omicas y Geof\'\i sicas, Universidad Nacional de La Plata, IALP-CONICET, Paseo del Bosque s/n, 1900 La Plata, Argentina
}


\altaffiltext{1}{{\it e-mail}: martinlc@iac.es}

\begin{abstract}
This is a brief note to comment on some recent papers addressing the Monoceros ring. In our view, nothing new was
delivered on the matter: No new evidence or arguments are presented which lead to think that the over-densities in Monoceros must not be due to the flared thick disc of the Milky Way. 

Again, we restate that extrapolations are easily misleading and that a model of the Galaxy is not the Galaxy. Raising and discussing exciting possibilities is healthy. However, enthusiasm should not overtake and produce strong claims before thoroughly checking simpler and more sensible possibilities within their uncertainties.
In particular, claiming that a reported structure, such as the Monoceros Ring, is not Galactic (an exciting scenario) should not be done without rejecting the possibility of being due to the well established warped and flared disc of the Milky Way (simpler).
\end{abstract}

\section{Discussion}

This is a brief note to comment on some recent papers addressing the Monoceros ring. As nothing new has been delivered on the matter, there is also nothing new to add. 
Concretely, the new data produce no evidence requiring the interpretation of the Monoceros over-density as a structure not belonging to the Milky Way.
Thus, there is no need to produce a full paper explaining why not. A simple note like the present one should be enough. Since silence gives consent---as the proverb goes---, here we break the silence and express our disagreement.

The so called Monoceros ring/stream is a hypothesis for explaining a reported over-density of stars (with respect to some standard models of the Milky Way, such as the Besan\c con model; Robin et al. 2003) in a
large area of the sky approximately parallel to the Galactic plane, in the latitude range $10^{\circ}<|b|<35^{\circ}$ and spanning most of the second and third quadrants (e.g., Ibata et al. 2003; Conn et al. 2008). It has been conjectured that this structure would be due to the remnants of a dwarf galaxy cannibalized by the Milky Way. The core of the progenitor would be associated to a further over-density of stars identified in the Canis Major subregion of the Monoceros ring (Martin et al. 2004).

However, the over-density of stars in Canis Major was soon explained
as an effect of the warped+flared disc of the Milky Way (Momany et al. 2004, 2006;
L\'opez-Corredoira 2006; L\'opez-Corredoira et al. 2007) with some features in the color-magnitude
diagrams due to the warped Norma--Cygnus spiral arm (Carraro et al. 2005, 2007, 2008; 
Moitinho et al. 2006; Piatti \& Clari\'a 2008). 
Since then, in the last years, the subject of the extragalactic origin of Canis Major was mostly dropped in the literature. It seemed that the purely Galactic explanation of the phenomenon had been mostly accepted.

Concerning the Monoceros ring as a whole, Ibata et al. (2003) and Momany et al. (2006) suggested that it can be explained by the flare of the Galactic outer disc, and Hammersley \& L\'opez-Corredoira (2011) made explicit calculations showing how a flare in the thick disc (an element not included in models such as the Besan\c con model) fits approximately the observed over-density in some regions of the anti-centre. 
The most recent deep surveys clearly show that there are stars out to at least $R$=20~kpc (e.g.,
Momany et al. 2006, Reyl\'e et al. 2009, Carraro et al. 2010). That the disc is flared, should not come as a surprise. It is expected (Momany et al. 2006) and not an ad-hoc hypothesis such as the one of an extragalactic stream (or a reported 
three-fold layer of streams; Li et al. 2012) parallel to the plane.
The explanations in terms of the structure of the Milky Way also contemplate the characteristics of the observed populations, including metallicities, kinematics and spatial densities. We find no observations leading to the necessity of considering the reported over-densities not naturally due to the structure the Milky Way.

Lately, four recent papers (Sollima et al. 2011; Meisner et al. 2012; Conn et al. 2012; Li et al. 2012) revive the claims of an extra-galactic origin for the Monoceros ring.
Below, we address the arguments and conclusions of these articles:
\begin{description}

\item[Density distribution:] Sollima et al. (2011) affirm that no model of the Milky Way is able to explain
the density distribution of the Monoceros structure. The statement is puzzling since Hammersley \& L\'opez-Corredoira (2011) had previously 
shown that a simple model of a flared thick disc does explain approximately the density distribution under discussion.
The lines of sight considered in both studies were close to each other [one of the lines of sight of Hammersley \& L\'opez-Corredoira (2011) is $\ell=183 ^\circ$, $b=21^\circ $, very close to the first field of Sollima et al. (2011) in $\ell=180^\circ $, $b=21^\circ $]. 
While Sollima et al. (2011) use the 
same model as Hammersley \& L\'opez-Corredoira (2011), Sollima et al. (2011) affirm that the model does not produce a detectable over-density bump. Figure 3 of Sollima et al. (2011) shows synthetic colour-magnitude diagrams with no clear main sequence. However, Hammersley \& L\'opez-Corredoira (2011) do reproduce such a main sequence in Monoceros at magnitudes between $g=20$ and 22, for a population of dwarfs with $g-r$ between 0.36 and 0.49. 
Moreover, other authors supporting the extra-Galactic scenario (e.g., Conn et al. 2012) could also reproduce the density distribution with a flared thick disc. Thus, we are led to impression that the calculations in Sollima et al. (2011) are in error and that a flared model can reproduce the over-density. 

Conn et al. (2012) also use the model of Hammersley \& L\'opez-Corredoira (2011) to fit the morphology of Monoceros over-density and arrive to conclusions
roughly similar to those of Hammersley \& L\'opez-Corredoira (2011). Although considering different regions than Hammersley \& L\'opez-Corredoira (2011) and achieving better fits with some different parameters with respect to Hammersley \& L\'opez-Corredoira (2011), Conn et al. (2012) arrive to qualitatively similar results. Interestingly, Conn et al. (2012) find that the over-density with respect to a non-flared model starts at distances of around 5-7 kpc from the Sun, whereas the regions closer to the anti-centre used by Hammersley \& L\'opez-Corredoira (2011) indicate a start at distances of 8-10 kpc. The difference is likely due to some lopsidedness of the disc or non-axisymmetry of the flare (L\'opez-Corredoira \& Betancort-Rijo 2009). The analysis of the density carried out by Li et al. (2012) is much simpler: the authors simply point out that the number of observed stars is much higher than that expected from a ``canonical'' thick disc, where by  ``canonical''  it is meant a non-flared model, but they cannot exclude a flared disc.

\item[Metallicity, stellar populations:] Although observations and discussions on the metallicity of Monoceros were already presented in many previous papers, Meisner et al. (2012), Conn et al. (2012), and Li et al. (2012) revive the theme with new data. These data yield the same results as before: $[Fe/H]\approx -1.0$ for the first two papers and $[Fe/H]\approx -0.8$ for Li et al. (2012). 
Surprisingly, and despite previous work (Momany et al. 2006, Hammersley \& L\'opez-Corredoira 2011), these new studies again affirm that such a metallicity is incompatible with the populations of our Galaxy.

Conn et al. (2012) argue that the thick disc should have an average metallicity $[Fe/H]\approx -0.6$ and no radial gradient. The argument is based on analyses of low $R$ and low $|z|$ stars extrapolated to high $R$ and high $|z|$. 
But extrapolations can easily be inadequate: the observed regions of Monoceros are 
at $R\approx 20$ kpc and $z\approx 4$ kpc for which there are no observations constraining the thick disc which are independent of Monoceros itself. 
As discussed below,  it is not surprising 
to find in this region stars with a metallicity 0.2-0.4 dex lower than stars at smaller $R$ and $|z|$. 
The statement by Conn et al. (2012) that there is no radial metallicity gradient
 in the thick disc is not strictly correct: As an example, in Cheng et al. (2012), which Conn et al. (2012) cite, it is shown that there is a significant negative gradient for stars with $[\alpha /Fe]<0.2$ (Cheng et al. 2012, Fig. 4), and Monoceros has indeed a significant number of stars with $[\alpha /Fe]<0.2$ (Meisner et al. 2012). Vertical gradients are also detected in the nearby thick disc (Bilir et al. 2012). 
In any case, a small average metallicity gradient of $\sim -0.03$ dex/kpc in the radial and vertical directions is enough to explain $[Fe/H]\approx -1.0$ and cannot be excluded at the present. 
Ironically, the Besan\c con model of the Milky Way --- used in arguing for an extragalactic origin of Monoceros in what concerns predicted stellar densities and colour-magnitude diagrams --- gives that the very outer thick disc should have an average metallicity $[Fe/H]$ very close to -1.0 (L\'opez-Corredoira et al. 2007, Fig. 3). 

Li et al. (2012) find that the Monoceros ring population at $g\approx 20$ has a bluer turn-off than the disc population which is closer at $g=17-18$. This is also expected since most Monoceros stars are presumably thick disc stars (with some small contribution from halo stars), whereas closer stars are a mixture of thin+thick disc with higher metallicity. The turnoff colour of $(g-r)=0.30-0.31$ measured by Li et al. (2012) for Monoceros is indeed very similar to the turnoff colour measured for the thick disc: $(g-r)\approx 0.33$ (Chen et al. 2001). There is no problem in interpreting this stellar population as belonging to the thick disc.
  
\item[Radial velocities:] Li et al. (2012) make an interesting point: that the line of sight velocities
in the range $150^\circ <\ell <190^\circ $ are significantly higher than those expected from a canonical
thick disc. In particular, that the average radial velocity at $\ell=180^\circ $ is significantly different from zero. As Li et al. (2012) state, all axisymmetric disc models predict a zero average line-of-sight velocity directly towards the anti-centre, independently of rotation speed and distance to the stars, because at that longitude we are looking perpendicularly to the circular motion of the disc stars. Li et al. (2012) interpret this non-zero detection as proof against a thick disc origin for Monoceros.
However, the assumption of perfectly circular orbits for the outer disc is not unquestionable. In fact, many spiral galaxies exhibit some non-axisymmetry/lopsidedness in their outer discs (Jog \& Combes 2009). There is also the possible phenomenon of stellar migration (Roskar et al. 2008) which displaces stars to different radii in non-circular orbits. A non-axisymmetric outer disc is not only a possibility, but it is also most likely. Thus, a non-zero average radial velocity towards the anti-centre is not enough to discard the thick disc origin of Monoceros.

\end{description}

We have shown how and why no new elements on the Monoceros affair are brought by a number of recent papers (Sollima et al. 2011, Conn et al. 2012, Meisner et al. 2012, and Li et al. 2012). We reaffirm that extrapolations are easily misleading and that {\bf a model of the Galaxy is not the Galaxy}. The Besan\c con model, despite being a very good model, is not correct in reproducing all the features of the outer disc. In particular, the Besan\c con model assumes a disc stellar truncation at $R\sim 14$ kpc.  
Although some studies argue for a cut-off of the stellar disc at that distance (e.g., Minniti et al. 2012), the deficit of outer in-plane stars is an expected artefact of assuming  a non-flared disc. Moreover, stars have been detected at $R\sim 20$ kpc with higher densities than those expected from models with a closer disc cut-off (e.g., Momany et al. 2006, Reyl\'e et al. 2009, Carraro et al. 2010). 
The unnecessary assumption of a close cut-off produces a cascade of further assumptions and loose ends of which the most dramatic is that the majority of stars at far galacto-centric distances are extragalactic.

Certainly, the hypothesis of Monoceros as an extragalactic tidal stream is not discarded, but there are no reasons to support it since it can be explained in terms of known features of our Galaxy. Such a strong claim should not be made without verifying the expected features, within the uncertainties, of the Galaxy. 
This note is a cautionary tale on how models should not be over-interpreted. Li et al. (2012) also discuss structures designated as the``Anti-Center Stream'' and the ``Eastern Banded Structure'' which do not look much better cases than Monoceros, but these will not be discussed here. 
Nonetheless, there are other tidal streams (e.g., Sagittarius) with solid observational support.

%{\bf Acknowledgments:}
%

%\bibliographystyle{apj}

\end{document}